\documentclass[12pt,preprint]{aastex}
\newcommand{\ltsimeq}{\raisebox{-0.6ex}{$\,\stackrel
        {\raisebox{-.2ex}{$\textstyle <$}}{\sim}\,$}}

\shorttitle{NGC\,6791}
\shortauthors{van Loon et al.}
\begin{document}
\title{{\it Spitzer} Space Telescope evidence in NGC\,6791: no super-mass-loss
at super-solar metallicity to explain helium white dwarfs?\altaffilmark{1}}
\author{Jacco Th. van Loon\altaffilmark{2},
Martha L. Boyer\altaffilmark{3},
Iain McDonald\altaffilmark{2}
}
\altaffiltext{1}{This work is based on observations made with the {\it Spitzer
Space Telescope}, which is operated by the Jet Propulsion Laboratory,
California Institute of Technology under a contract with NASA.}
\altaffiltext{2}{Astrophysics Group, Lennard-Jones Laboratories, Keele
University, Staffordshire, ST5 5BG, UK; jacco@astro.keele.ac.uk,
iain@astro.keele.ac.uk.}
\altaffiltext{3}{Dept.\ of Astronomy, School of Physics \& Astronomy, 116
Church Street S.E., University of Minnesota, Minneapolis, MN 55455;
mboyer@astro.umn.edu.}
\begin{abstract}
We use archival {\it Spitzer} Space Telescope photometry of the old,
super-solar metallicity massive open cluster NGC\,6791 to look for evidence of
enhanced mass loss, which has been postulated to explain the optical
luminosity function and low white dwarf masses in this benchmark cluster. We
find a conspicuous lack of evidence for prolificacy of circumstellar dust
production that would have been expected to accompany such mass loss. We also
construct the optical and infrared luminosity functions, and demonstrate that
these fully agree with theoretical expectations. We thus conclude that there
is no evidence for the mass loss of super-solar metallicity red giants to be
sufficiently high that they can avoid the helium flash at the tip of the red
giant branch.
\end{abstract}
\section{Introduction}

Low-mass stars in galactic globular clusters lose $\sim20$\% of their mass
during the ascent of the first red giant branch
\citep[RGB;][]{Dupree07,Origlia07,Mcdonald07}, offering a partial explanation
for the blue extent of the core-helium burning horizontal branch
\citep{Rood73}. It has been argued recently that in the old (8 Gyr),
super-solar metallicity ($Z\simeq0.05$) massive open cluster NGC\,6791, at
least 40\% of RGB stars experience higher rates of mass loss; this would lead
to their departure from the RGB before the helium flash, resulting in
undermassive helium white dwarfs \citep{Castellani93,Hansen05,Kalirai07}. If
true, this would have important implications for the white dwarf populations
produced in metal-rich stellar systems, and for the mass return into the
interstellar medium of such systems --- which include giant elliptical
galaxies and the bulges of spiral galaxies. \citet{Bedin08}, on the other
hand, propose an alternative explanation for the white dwarf properties in
NGC\,6791 without the need for enhanced RGB mass loss.

We investigate the evidence for super-mass-loss on the RGB of NGC\,6791 by
looking for circumstellar dust that may form in the winds from cool RGB stars
\citep[e.g.,][]{Gehrz71,Origlia07} in archival {\it Spitzer} Space Telescope
infrared (IR) images, and by comparing the IR and optical luminosity functions
with model predictions.

\section{{\it Spitzer} photometry of NGC\,6791}

\begin{figure}
\plotone{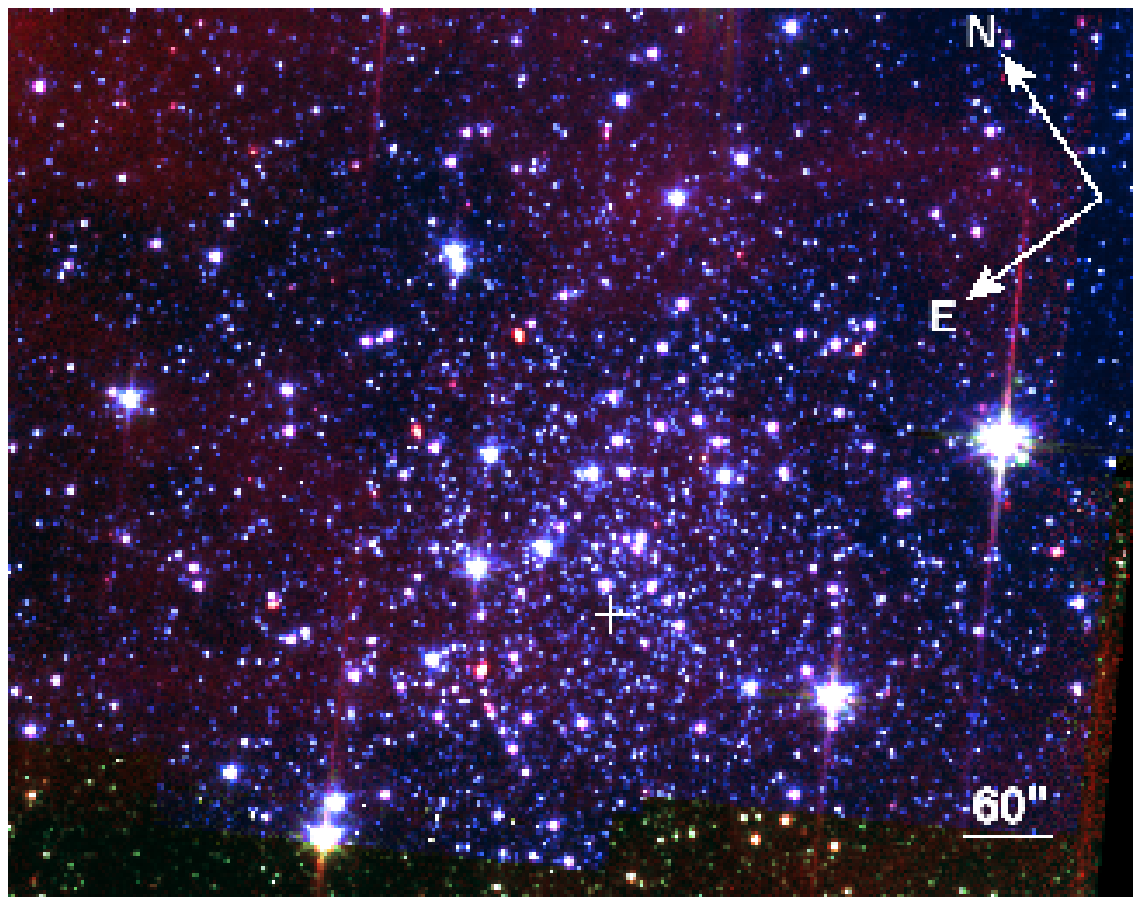}
\caption{{\it Spitzer} IRAC composite of the region around NGC\,6791
(indicated with the cross). Red, yellow-green, green-cyan, and blue are 8,
5.8, 4.5, and 3.6 $\mu$m, respectively.}
\label{f1}
\end{figure}

The observations of NGC\,6791 presented here were intended for {\it James
Webb} Space Telescope photometric and astrometric calibration
\citep{Diazmiller07}; we obtained the data from the {\it Spitzer} Space
Telescope \citep[hereafter {\it Spitzer};][]{Werner04} public archive. The
observations were made with the {\it Spitzer} Infrared Array Camera
\citep[IRAC;][]{Fazio04} on 2005 June 14 UT.  The image maps cover
$\approx20^\prime\times15^\prime$, with 3.6 and 5.8 $\mu$m mosaics offset to
the northwest from the cluster center by $\approx8^\prime$ and by
$\approx3^\prime$ at 4.5 and 8 $\mu$m. The region around the cluster is
displayed in Fig.\ \ref{f1}. High dynamic range mode had been implemented to
prevent saturation of brighter sources (1 s exposures) while maintaining a
high signal-to-noise on fainter sources (26.8 s exposures), and observations
had been made with an eight position cycling dither pattern to build
redundancy against outliers and artefacts.

Raw data were processed with {\it Spitzer} Science Center (SSC) pipeline
version S14.0.0. We corrected the Basic Calibrated Data images for array
distortions and mosaicked the resulting images with the SSC Legacy MOPEX
software v16.2.1 \citep{Makovoz05}, which includes background-matching to
minimize pixel offsets in overlapping regions of the mosaics and eliminates
cosmic rays and other outliers.

We carried out point-spread function photometry using the DAOphot II
photometry package \citep{Stetson87}. Sources at least 4$\sigma$ above the
background were extracted. A sharpness cut-off helped eliminate extended
sources and remaining outliers. A pixel-phase-dependent photometric correction
was applied to the 3.6 $\mu$m fluxes \citep{Reach05}. The flux estimates are
color corrected using a 5000 K blackbody, as specified in the IRAC Data
Handbook, version 3.0. Zero magnitudes were taken from the same reference, and
flux uncertainties include both the DAOphot errors and calibration errors
\citep{Reach05}. False star tests indicate that the catalogue is complete to
at least 90\% in the core of the cluster down to $[3.6]=17.5$ mag and
$[8]=16.5$ mag.

\section{Cluster membership}

\begin{figure}
\plotone{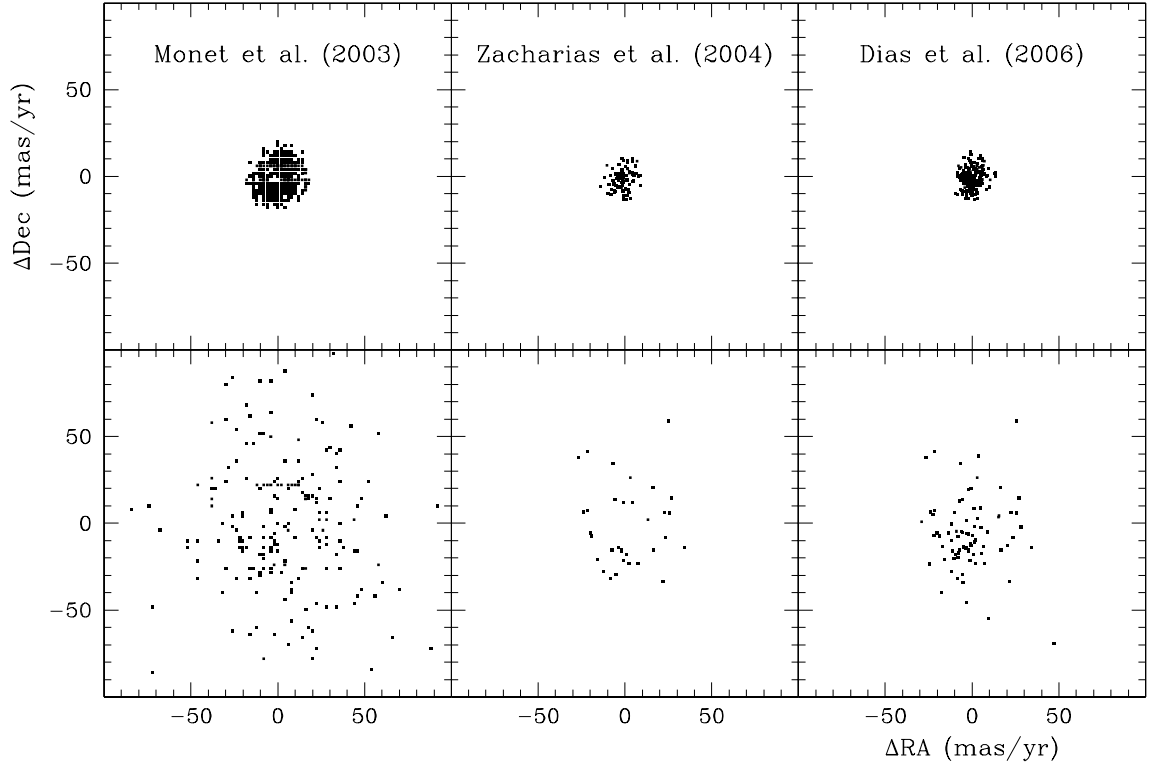}
\caption{Proper motion maps based on data from \citet{Monet03},
\citet{Zacharias04} and \citet{Dias06}, which are used to select against
non-members of NGC\,6791 (bottom row of panels).}
\label{f2}
\end{figure}

Proper motions have been measured for many stars in and around NGC\,6791
\citep{Monet03,Zacharias04,Dias06}. We designate a star a proper motion
non-member if the absolute proper motion is $>20$ mas yr$^{-1}$ in the
\citet{Monet03} catalogue, $>14$ mas yr$^{-1}$ in the \citet{Zacharias04}
catalogue, or its membership probability was set zero in the \citet{Dias06}
catalogue. Fig.\ \ref{f2} displays the resulting selection of possible members
(top row) and proper motion non-members (bottom row). The figure illustrates
the choice of selection criteria for the \citet{Monet03} and
\citet{Zacharias04} catalogues, which are based on the clearly defined clump
of stars with small proper motions (possible members) compared to the much
more widely distributed foreground population. The \cite{Dias06} selection
criteria include a positional element. Note that if a star was excluded as a
member by any one of these works it is labelled a non-member in all three
panels of Fig.\ \ref{f2}.

\begin{figure}
\plotone{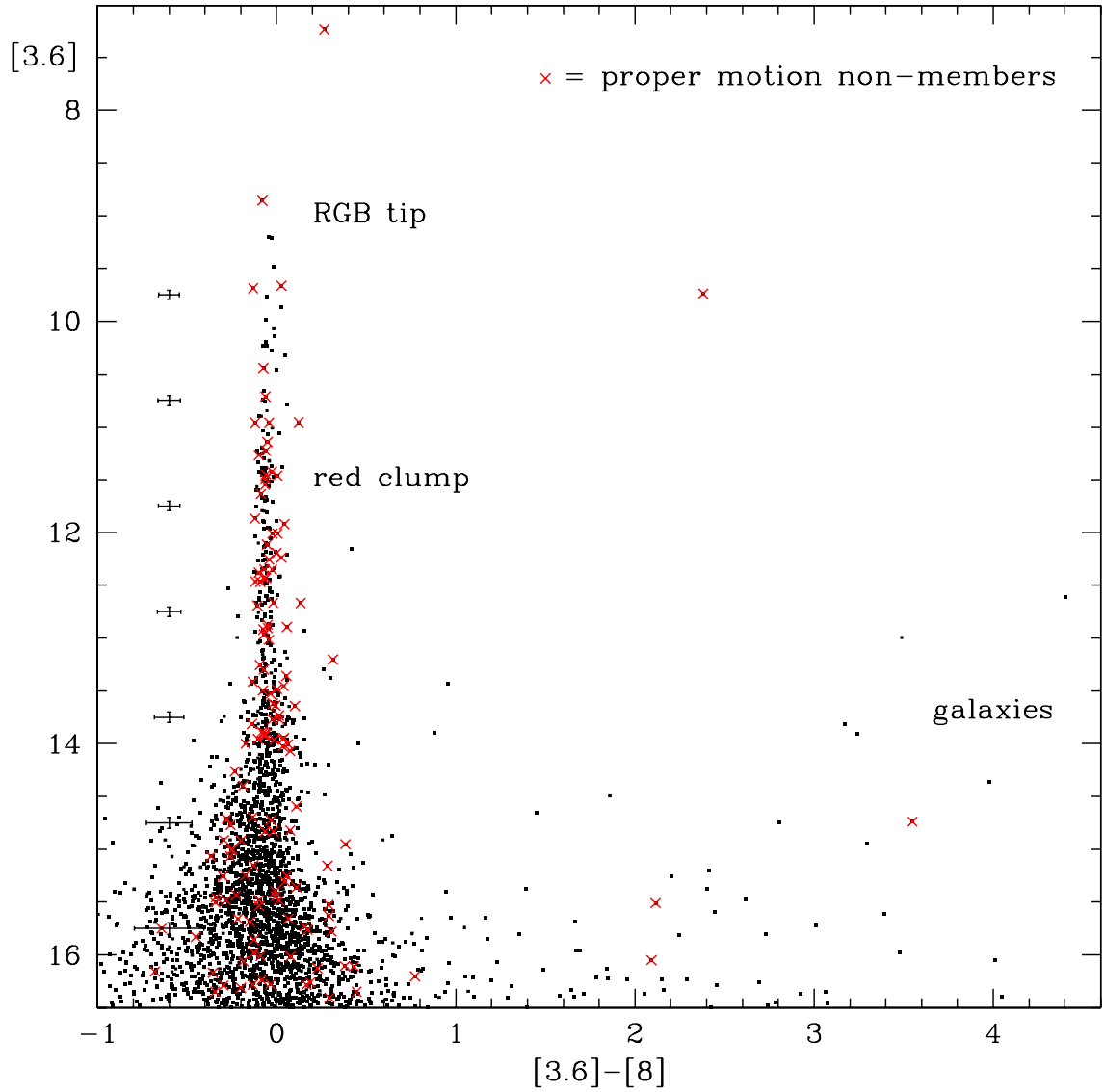}
\caption{{\it Spitzer} IRAC color-magnitude diagram of NGC\,6791. The RGB is
well populated until the tip at [3.6]$\simeq9$ mag, and the red clump is
visible around [3.6]=11.5 mag. Hardly any star shows evidence for reddening
due to a contribution of circumstellar dust emission at 8 $\mu$m.}
\label{f3}
\end{figure}

The proper motion non-members are identified with crosses in the {\it Spitzer}
[3.6], [3.6]--[8] color-magnitude diagram (Fig.\ \ref{f3}). This eliminates
for instance the very brightest star \citep[\#2240 in][]{Stetson03} and the
bright very red star at [3.6]=9.7 mag, [3.6]--[8]=2.4 mag \citep[\#2349
in][]{Stetson03}. Objects in the faint, red corner of the diagram are
background galaxies; slightly extended objects with similar colors are
prominent on the {\it Spitzer} images. As expected for galaxies, these objects
are spread uniformly across the images.

\begin{figure}
\plotone{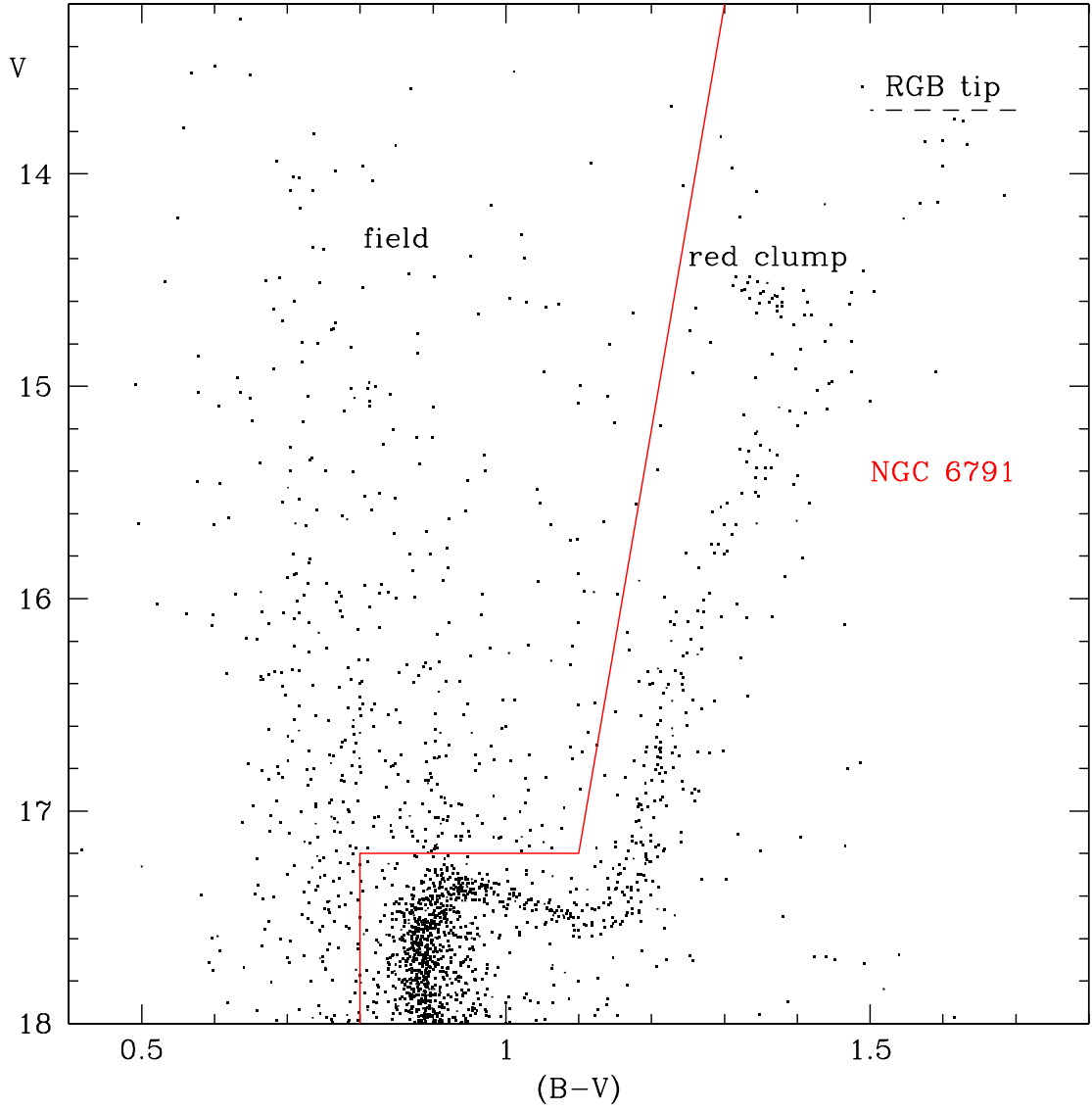}
\caption{Optical color-magnitude diagram \citep{Stetson03}, used to exclude
remaining field stars left of the dividing line.}
\label{f4}
\end{figure}

To eliminate field stars with no or ambiguous proper motion information, a
further criterion is applied on the basis of the optical V, B--V diagram
\citep[Fig.\ \ref{f4}, using data from][]{Stetson03}: the main sequence, RGB
and red clump (around $V=14.6$ mag, $B-V\simeq1.35$ mag) of NGC\,6791 are well
separated from the bulk of the foreground main-sequence stars. In our
subsequent analysis we keep all remaining possible cluster members that have
proper motion information, photometry in all {\it Spitzer} IRAC bands and the
V-band, and that are located within $0.1^\circ$ projected distance from the
cluster centre (290.23, +37.77).

\section{Evidence for circumstellar dust emission}

\begin{figure}
\plotone{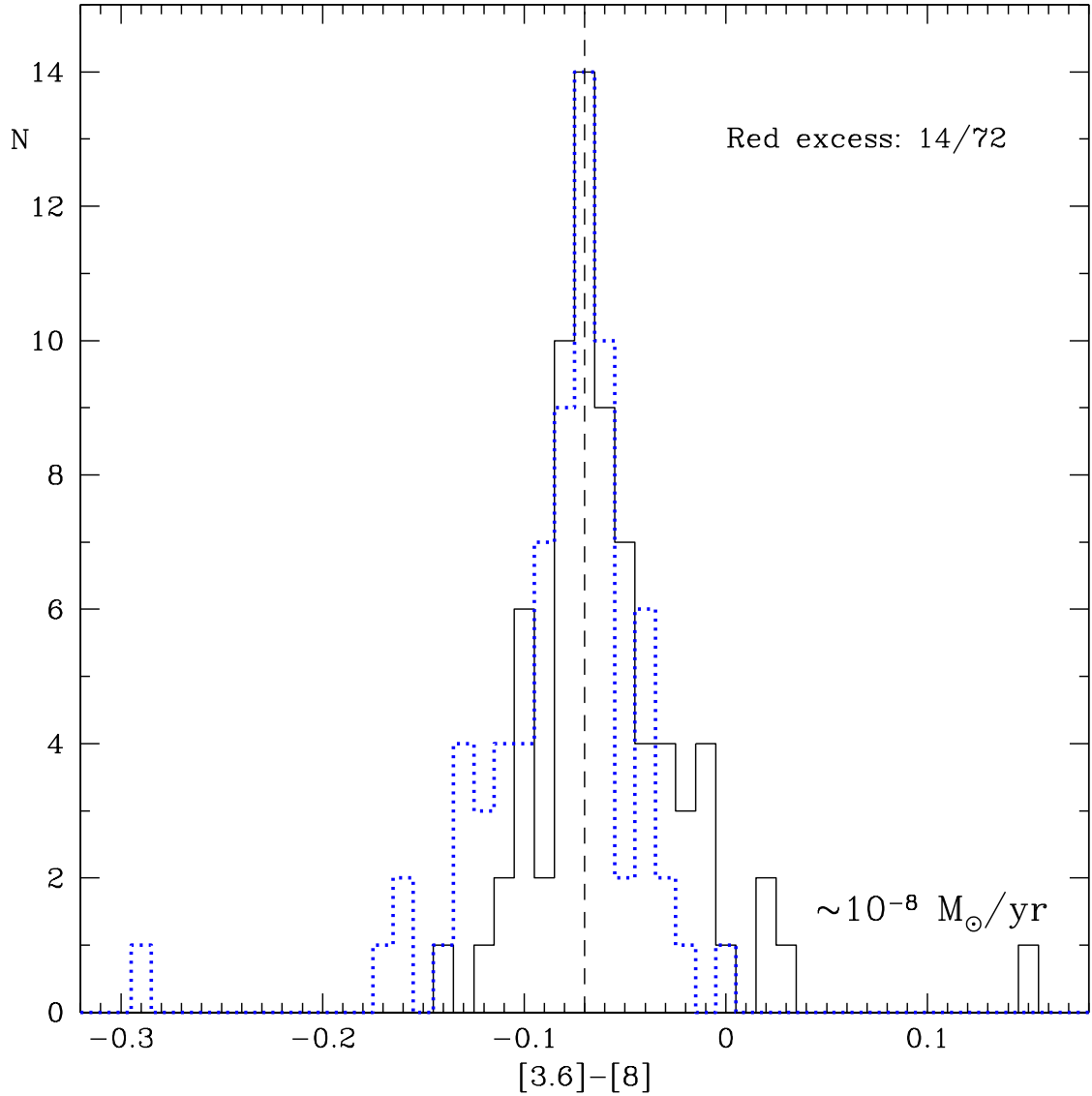}
\caption{{\it Spitzer} IRAC color distribution along the RGB in NGC\,6791.
Overplotted (dotted) is the same histogram, but mirrored with respect to its
peak. This more clearly shows which stars are redder than the bulk of RGB
stars.}
\label{f5}
\end{figure}

In the absence of longer-wavelength data, [3.6]--[8] is the most sensitive
{\it Spitzer} color to the presence of circumstellar dust. Dust causes a
reddening of this color, firstly, due to emission at 8 $\mu$m, and secondly,
due to extinction at 3.6 $\mu$m if the dust envelope is very optically thick.
The RGB in NGC\,6791 is surprisingly narrow in [3.6]--[8] color (Fig.\
\ref{f3}) down to at least $[3.6]\simeq13.5$ mag, i.e.\ more than four
magnitudes below the RGB tip. Compared to the distribution mirrored around the
peak (Fig.\ \ref{f5}), at $[3.6]-[8]=-0.07$ mag, there are 14 out of 72 stars
($19\pm5$\%) that are redder than expected from a distribution which is
symmetric in color. Not only is this but a small fraction, the colors are also
only marginally red, with $\Delta([3.6]-[8])<0.1$ mag for all but one with
$\Delta([3.6]-[8])\simeq0.22$ mag. The latter is at the faint end of this RGB
selection; indeed, there is no sign of the brightest stars being more likely
to show an IR excess as it is observed, e.g., in $\omega$\,Centauri
\citep{Boyer08}.

The fraction of dusty stars along the upper 4.5 mag of the RGB in NGC\,6791
may be compared with that in the globular cluster 47\,Tuc \citep{Origlia07}.
There, $\sim340$ stars are counted in the central part of 47\,Tuc, amongst
which 90 are dusty (26\%). In the outskirts of 47\,Tuc, there are $\sim160$
stars, amongst which 9 are dusty (6\%). These statistics may be affected by
blending, in the central part, and field star contamination, in the outskirts.
Nonetheless, there is no evidence for an abnormally high fraction of dusty
stars in NGC\,6791 despite an order of magnitude higher metal content than in
47\,Tuc.

Following \citet{Groenewegen06}, $\Delta([3.6]-[8])=0.1$ mag corresponds to a
mass-loss rate of $1-2\times10^{-8}$ M$_\odot$ yr$^{-1}$ for an M-type star
with aluminium-oxide or silicate dust and a luminosity $L=3,000$ L$_\odot$,
wind speed $v=10$ km s$^{-1}$, and dust-to-gas mass ratio $\psi=0.005$. In the
extreme case of an M10-type star with pure aluminium-oxide dust the
corresponding mass-loss rate is nearly $5\times10^{-8}$ M$_\odot$ yr$^{-1}$,
but to our knowledge no such RGB star exists. The mass-loss rate scales
approximately as $\dot{M}\propto v \sqrt{L}/\psi$ \citep{Vanloon00}. For a
dust-driven wind one expects the dust-to-gas ratio in the metal-rich stars of
NGC\,6791 to be higher by a factor $\sim2.5$, and the wind speed by a factor
$\sim1.6$ \citep{Marshall04}. The luminosity range along this part of the RGB
is $\sim10^{2-3}$ L$_\odot$, which lowers the Groenewegen value for $\dot{M}$
by up to a factor 5. If the winds are as fast as sometimes seen in
chromospherical lines of warmer RGB stars \citep{Mauas06,Mcdonald07} then
$\dot{M}$ could be higher. We thus settle on a conservative estimate of
$\dot{M}\ltsimeq10^{-8}$ M$_\odot$ yr$^{-1}$. The mass-loss rate may still be
higher if the dust-to-gas ratio is lower than assumed --- in spite of the high
abundance of condensable material. However, our estimate compares favorably
with the empirical relationships of \citet{Reimers77} and \citet{Schroeder05},
which yield $\dot{M}\sim10^{-8}$ and a few $10^{-9}$ M$_\odot$ yr$^{-1}$,
respectively, for a typical RGB star of 1 M$_\odot$, 4000 K and 500 L$_\odot$.
This suggests that the basic assumptions we made are not unreasonable.

If such mass loss is sustained over 19\% (the observed fraction of dusty
stars) of the time it takes to evolve along this part of the RGB,
$t\sim8\times10^7$ yr \citep{Marigo08}, then they will lose
$\Delta M\ltsimeq0.2$ M$_\odot$ --- insufficient to avoid the helium flash.
These stars were born with a mass $M_{\rm initial}\simeq1.1$ M$_\odot$, and
would need to shed $\Delta M_{\rm implied}\simeq0.7$ M$_\odot$ to produce
white dwarfs of the masses measured by \citet{Kalirai07}. Both
\citet{Kalirai07} and \citet{Hansen05} point out that not all RGB stars may
avoid the helium flash. Given the above estimates of $\dot{M}$, fraction of
dusty stars and RGB lifetime, the {\it Spitzer} data suggest that no more than
$\dot{M} t/\Delta M_{\rm implied} \times 19$\% or $\ltsimeq22$\% of RGB stars
in NGC\,6791 can sustain the required heavy mass loss.

There is a possibility that we have missed a star at the very tip of the RGB
losing mass at a high rate, but only briefly before it leaves the RGB. Indeed,
dust-accompanied mass loss, at rates as high as $10^{-6}$ M$_\odot$ yr$^{-1}$,
is seen to happen predominantly near the tip of the RGB
\citep[][]{Vanloon06,Origlia07,Boyer08}. It is unclear, however, whether such
star has sufficient time to lose enough mass to avoid the helium flash. In any
case, despite the important effect on the remaining mantle mass and its
subsequent appearance on the horizontal branch, the white dwarf descendant
would have a mass that is indistinguishable from that of a CO core following
the helium flash, as the luminosity on the RGB is a direct measure of the
helium core's mass \citep{Castellani93}.

\section{The RGB luminosity function}

To explain white dwarf masses 0.1 M$_\odot$ lower than the core mass at the
tip of the RGB, the luminosity function must be depleted over 0.6 dex in $\log
{\rm L}$ \citep[Table 2 in][]{Castellani93}, corresponding to 1.5 mag. We
demonstrate below that such depletion in NGC\,6791 is not corroborated by the
measurements.

\begin{figure}
\plotone{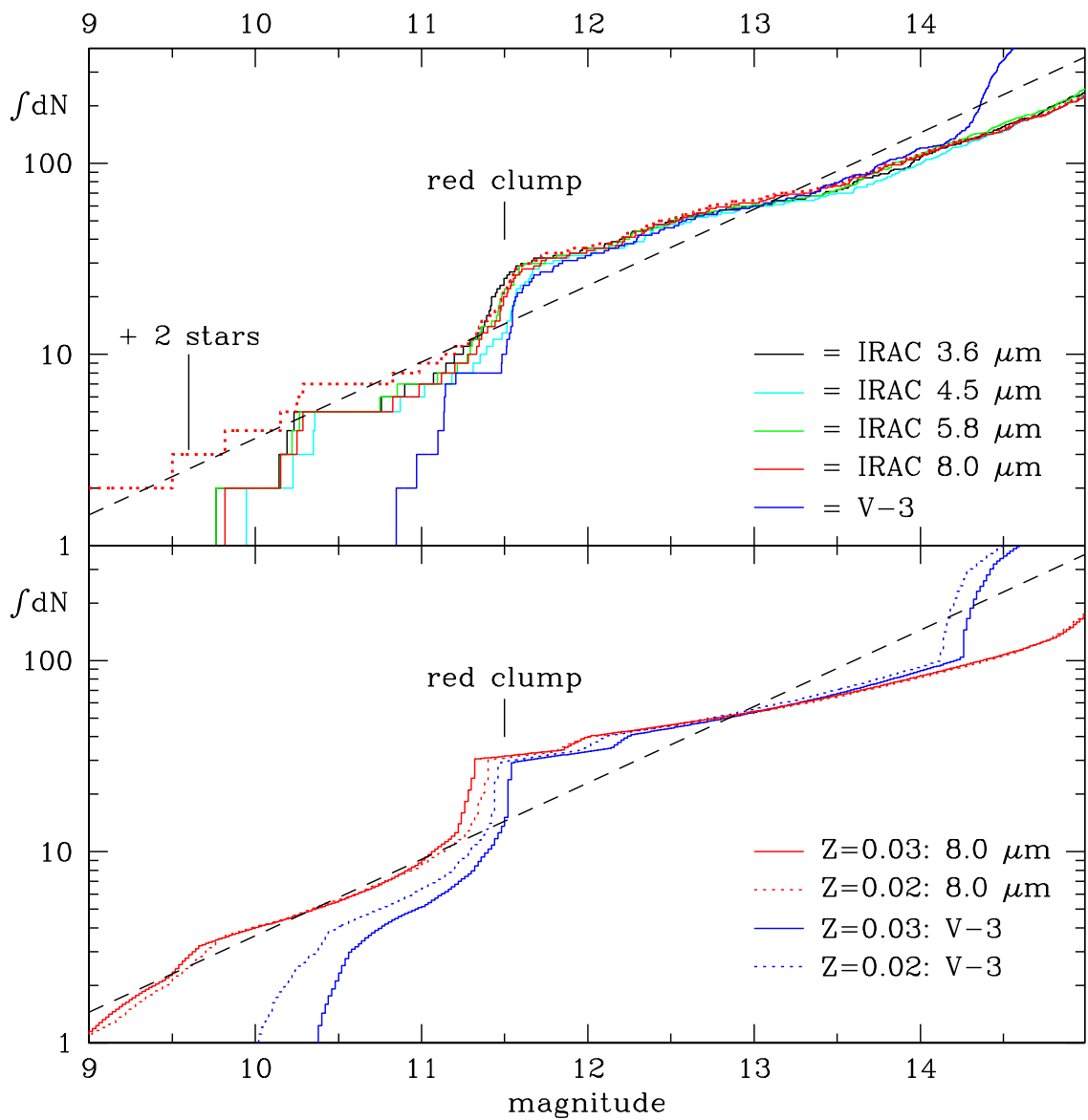}
\caption{{\it Spitzer} IRAC and optical V-band \citep{Stetson03} luminosity
functions in NGC\,6791. The 8-$\mu$m luminosity function is also plotted after
adding just two more stars near the tip of the RGB. A comparison with 8-Gyr
Padua models \citep{Marigo08} shows good agreement, especially considering
that NGC\,6791 is more metal-rich. Stochastics introduced by sampling the
small number of stars on the upper RGB can fully explain the discrepancy
between the data and models.}
\label{f6}
\end{figure}

We constructed cumulative luminosity functions (LFs), in both the {\it
Spitzer} IRAC bands and the V-band (Fig.\ \ref{f6}). The IR and V-band LFs
were aligned such that the very obvious bump due to the red clump coincides
--- this required a shift towards brighter values of the V-band LF by 3 mag.
This difference, and the much steeper bright end to the V-band LF compared to
the IR LFs, are due mainly to the bolometric corrections to the V-band
(BC$_{\rm V}$) of the increasingly cooler RGB stars as they approach the RGB
tip \citep[a small contribution is due to the extinction,
$E(B-V)\simeq0.14$;][]{Kalirai07}.

The observed LFs compare favorably with model LFs from the Padua group
\citep{Marigo08}, displayed in the lower panel of Fig.\ \ref{f6}. The models,
displayed for a distance modulus of 13.0 mag, show the same shape and location
of the red clump and main sequence (the rise in the V-band LF at the faint end
of the scale). The models, which do not include mass loss along the RGB,
reproduce the observed difference at the bright end between the IR and V-band
LFs rather well. The most metal-rich Padua model is with $Z=0.03$ not quite as
metal-rich as NGC\,6791 ($Z=0.05$), and if compared to the $Z=0.02$ Padua
model it is evident that a $Z=0.05$ model LF would be truncated at even
fainter magnitudes, in close agreement with the observed LF. This is
corroborated by the good fit to the optical colour-magnitude diagram of a
$Z=0.06$ isochrone using the \citet{Reimers77} mass-loss prescription
\citep{Claret07}. \citet{Kalirai07} find that the V-band LF of NGC\,6791 is
depleted with respect to the (sub-)solar metallicity clusters Berkeley\,17 and
M\,67, over the top magnitude of the RGB. They do not, however, consider that
BC$_{\rm V}$ also differs by $>1$ mag \citep{Marigo08}, nor the effects from
contamination by field stars.

With regard to the IR LF, which more accurately measures the bolometric output
of these stars, the difference with the models can be entirely explained by
stochastic effects. The NGC\,6791 cluster is not populous enough to harbor
many stars near the RGB tip; while there are few tip-RGB stars present now, it
is conceivable that one or two more may have been present several million
years ago or will be present several million years from now. It can also not
be excluded that a tip-RGB star was missed, for instance, because it had
fallen outside the {\it Spitzer} coverage, or because of an inaccuracy in the
proper motion measurement, or for other observational reasons. To demonstrate
the feebleness of any indication of a prematurely truncated RGB, the 8-$\mu$m
LF is shown again after adding a mere two stars at the tip of the RGB. This
brings the observed LF in excellent agreement with the theoretical prediction.
Indeed, the relative numbers of post-helium-flash red clump stars compared to
those at similar luminosities on the RGB are well reproduced (Fig.\ \ref{f6}),
suggesting that the vast majority of RGB stars do go on to undergo the helium
flash.

\section{Conclusions}

We conclude that there is good agreement between the observed and expected
optical and infrared luminosity functions along the RGB in NGC\,6791, and that
there is little circumstellar dust observed around RGB stars in that old
metal-rich massive open cluster. Hence there is no direct evidence supporting
the suggestion that metal-rich stars avoid the helium flash at the tip of the
RGB and become undermassive helium-core white dwarfs as a result of
particularly strong stellar winds.

\acknowledgments

We thank both referees for their thorough reviews. MLB is supported by the
University of Minnesota Louise T.\ Dosdall Fellowship. IM acknowledges an STFC
studentship.


\clearpage

\end{document}